\documentclass{PoS}
\usepackage{caption}
\usepackage{subcaption}
\usepackage{graphicx}
\usepackage{nicefrac}
\usepackage{citesort}
\usepackage{wrapfig}

\usepackage{caption,setspace}
\usepackage{lipsum,lineno}
\usepackage[utf8]{inputenc}
\usepackage[none]{hyphenat}
\makeatletter 
\renewcommand\speaker[1]{\if@speaker\global\@dblspeaktrue\fi
			\global\@speakertrue
			\global\setbox\@firstaubox
			\hbox{{\let\thanks\@gobble
				\let\footnote\@gobble\small
				\rm  The xFitter Developers' Team}}%
			#1\thanks{Speaker.}\
			}%
\makeatother

\title{xFitter 2.0.0: An  Open Source QCD Fit Framework\thanks{%
      We would like to acknowledge Ringaile Pla\v cakyt\.e who served as
convener of the HERA-Fitter/xFitter project from 2012 until March 2017,
and Voica Radescu who served as founder and convener from the start
of the project in 2011 until June 2017. The outstanding dedication
and service of these colleagues to this project was immeasurable and
invaluable.
\newline
\indent
We acknowledge the hospitality of CERN, DESY, and Fermilab where a
portion of this work was performed.
This work was also partially supported by the U.S.\ Department of
Energy under Grant No.\ DE-SC0010129. 
We are grateful to the DESY IT department for their support of the
xFitter developers.
}}

\ShortTitle{xFitter 2.0.0: An  Open Source QCD Fit Framework}

\def\thanksref#1{\rlap,${}^{#1}$}
\def\inst#1{\hangafter=1\hangindent=15pt\relax ${}^{#1}$}

\author{
The xFitter Developers' Team: \quad
V.~Bertone\thanksref{a,b} \  
M. Botje\thanksref{c} \
D. Britzger\thanksref{d} \
S.~Camarda\thanksref{e}   \
A.~Cooper-Sarkar\thanksref{f} \
F.~Giuli\thanksref{f} \
A.~Glazov\thanksref{d} \
A.~Luszczak\thanksref{g} \  
F.~Olness\thanksref{h}\speaker{} \
R.~Pla\v cakyt\.e\thanksref{i} \
V.~Radescu\thanksref{e,f}  \
W.~S\l{}omi\'nski\thanksref{j} \ 
and \
O.~Zenaiev${}^{d}$ 
\\
\inst{a} Department of Physics and Astronomy,  VU University, NL-1081 HV Amsterdam, The~Netherlands\\
\inst{b} Nikhef Theory Group Science Park 105, 1098 XG Amsterdam, The Netherlands \\
\inst{c} Nikhef, Science Park, Amsterdam, The Netherlands\\
\inst{d} DESY Hamburg, Notkestra{\ss}e 85, D-22609, Hamburg, Germany \\
\inst{e} CERN, CH-1211 Geneva 23, Switzerland\\
\inst{f} University of Oxford,1 Keble Road, Oxford OX1 3NP, United Kingdom \\
\inst{g} T.Kosciuszko Cracow University of Technology, 30-084 Cracow, Poland \\
\inst{h} SMU Physics, Box 0175 Dallas, TX  75275-0175, United States of America \\
\inst{i} Institut f\"ur Theoretische Physik, Universit\"at Hamburg,  Luruper Chaussee 149, \linebreak[1]  D--22761 Hamburg, Germany \\
\inst{j} M. Smoluchowski Institute of Physics, Jagiellonian University, \L{}ojasiewicza 11, \linebreak[1]  30-348 Krak\'ow, Poland \\
}

\abstract{%
xFitter~\cite{Alekhin:2014irh}  is an open-source package that provides a framework for the
determination of the parton distribution functions (PDFs) of the
proton  for many different kinds of analyses in Quantum
Chromodynamics (QCD). 
xFitter version 2.0.0 has recently been released, and offers an expanded set of tools and options. 
It incorporates experimental data from a wide range of experiments 
including fixed-target, Tevatron, HERA, and LHC. 
xFitter can analyze this data up to 
next-to-next-to-leading-order (NNLO)  in perturbation theory with a variety of 
theoretical calculations including   numerous
methodological options for carrying out PDF fits and plotting tools which
help visualize the results. 
While primarily based on the collinear factorization foundation, 
xFitter also provides facilities for fits of dipole models and transverse-momentum dependent (TMD) PDFs. 
The package can be used to study the impact of new precise measurements from
hadron colliders, and  also assess the  impact of future  colliders.  
This paper provides a brief overview of xFitter 
with emphasis of the new version 2.0.0 features. 
}

\FullConference{XXV International Workshop on Deep-Inelastic Scattering and Related Subjects\\
		3-7 April 2017\\
		University of Birmingham, UK}

\begin{document}

% FIG 1
%%%%%%%%%%%%%%%%%%%%%%%%%%%%%%%%%%%%%%%%%%%%%%%%%%%%%%%%%%%%
\begin{figure}[th]
\vspace{-0.3cm}
  \centering
    \includegraphics[width=0.90\textwidth]{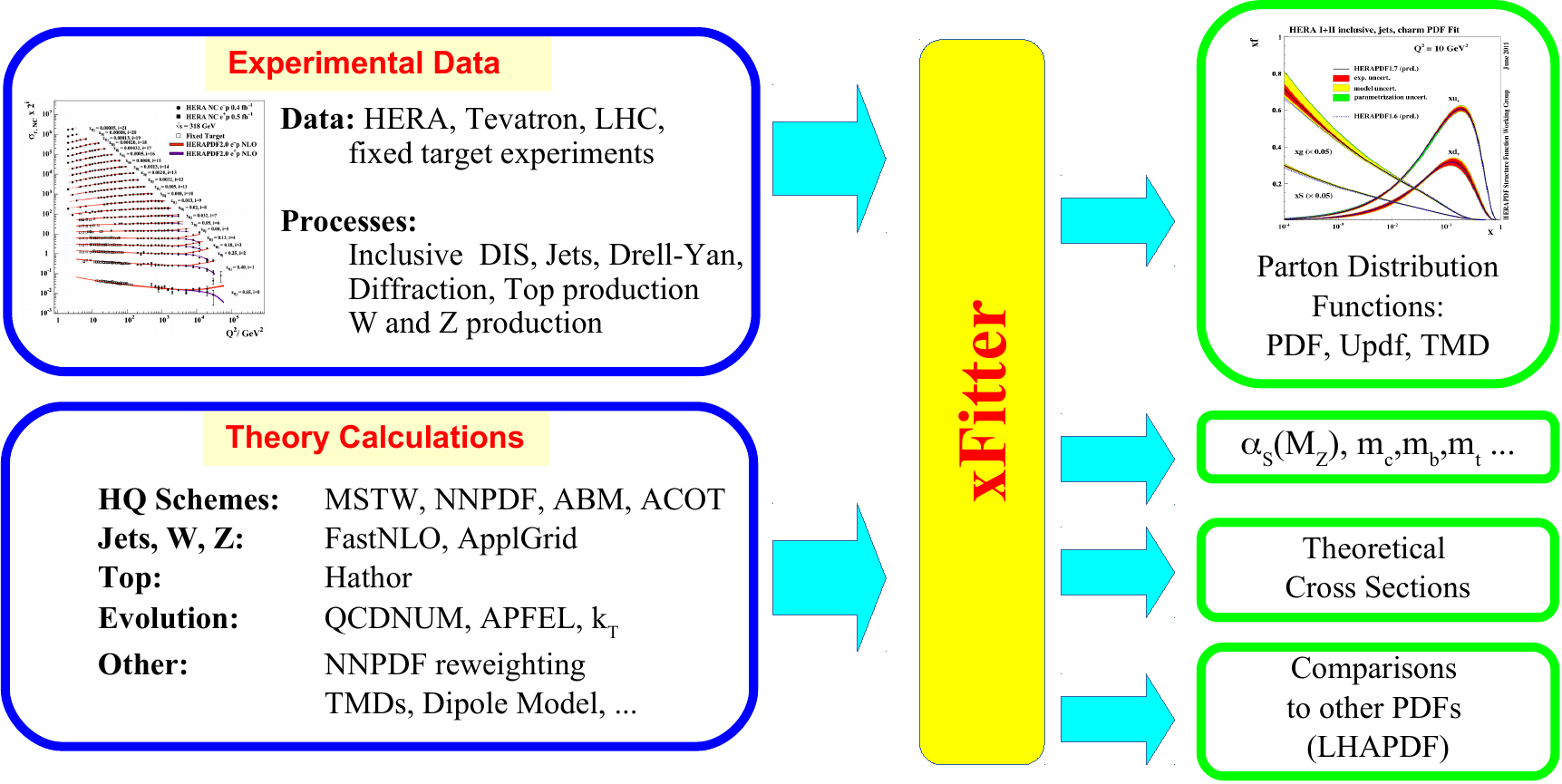}
    \caption{Schematic  of the modular structure of  xFitter  
illustrating the components and capabilities of the program.  \label{fig:flow}
}
\vspace{-0.8cm}
\end{figure}
%%%%%%%%%%%%%%%%%%%%%%%%%%%%%%%%%%%%%%%%%%%%%%%%%%%%%%%%%%%%

%%%%%%%%%%%%%%%%%%%%%%%%%%%%%%%%%%%%%%%%%%%%%%%%%%%%%%%%%%%%
\section{Introduction}

%FIG FROZEN FROG
%%%%%%%%%%%%%%%%%%%%%%%%%%%%%%%%%%%%%%%
\begin{wrapfigure}{R}{2.9cm} %%%%%%%%%%%%%%%%%%%%%
\centering{} 
\vspace{-40pt}
\includegraphics[width=2.9cm]{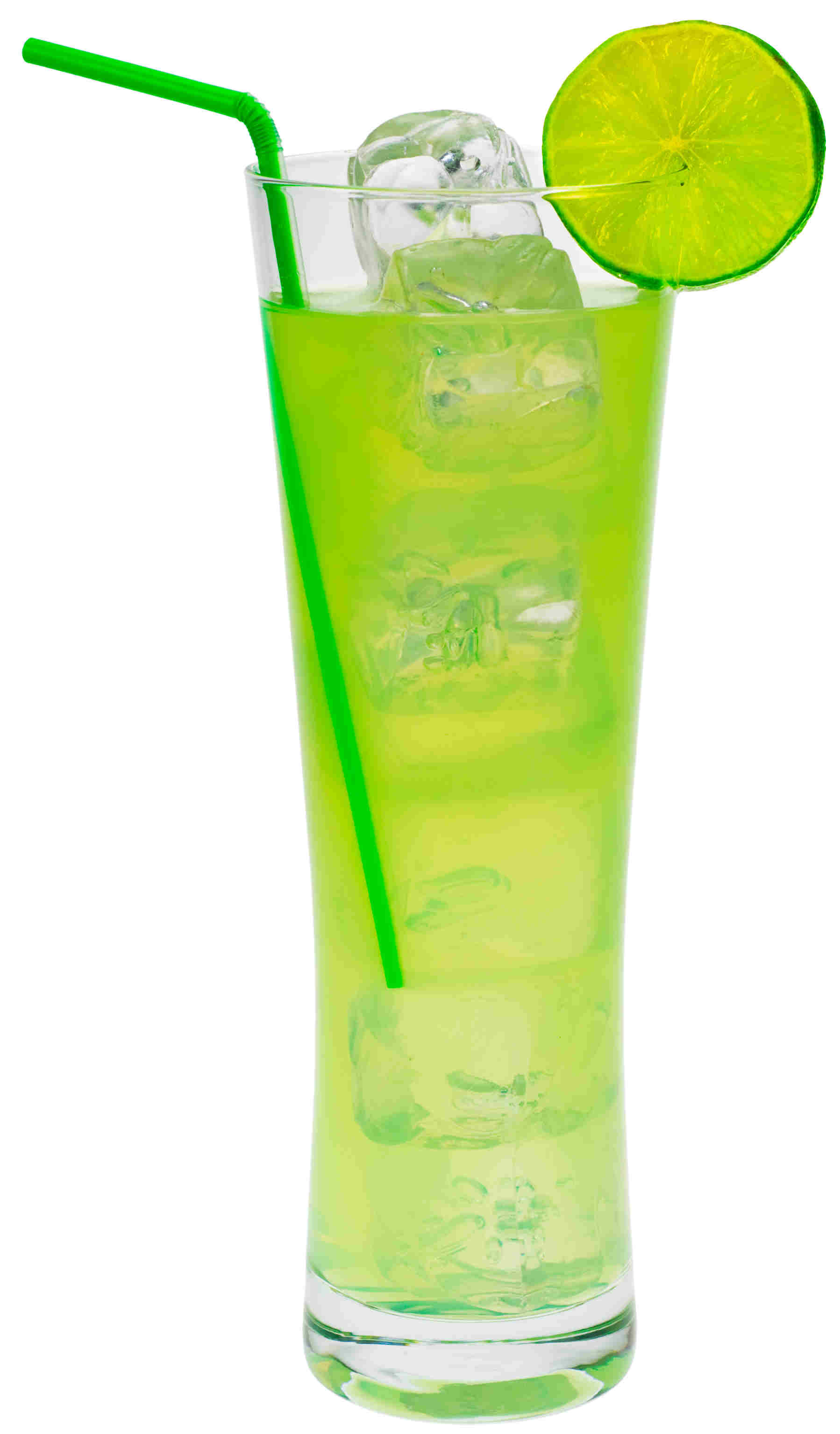}
\vspace{-20pt}
\caption*{
\linespread{0.9}\selectfont{}
\centering{} 
\null
\centerline{xFitter~2.0.0 \quad} 
\centerline{\it (Frozen~Frog)}
{\tiny iStock.com/Enjoylife2}
\vspace{-10pt}
}
\label{fig:one}
\end{wrapfigure}
%%%%%%%%%%%%%%%%%%%%%%%%%%%%%%%%%%%%%%%%%%%%

%FIG 2,3
%%%%%%%%%%%%%%%%%%%%%%%%%%%%%%%%%%%%%%%%%%%%%%%%%%%%%%%%%%%%
\begin{figure}[t]
  \centering
  \begin{minipage}[t]{0.45\textwidth}
    \includegraphics[width=\textwidth]{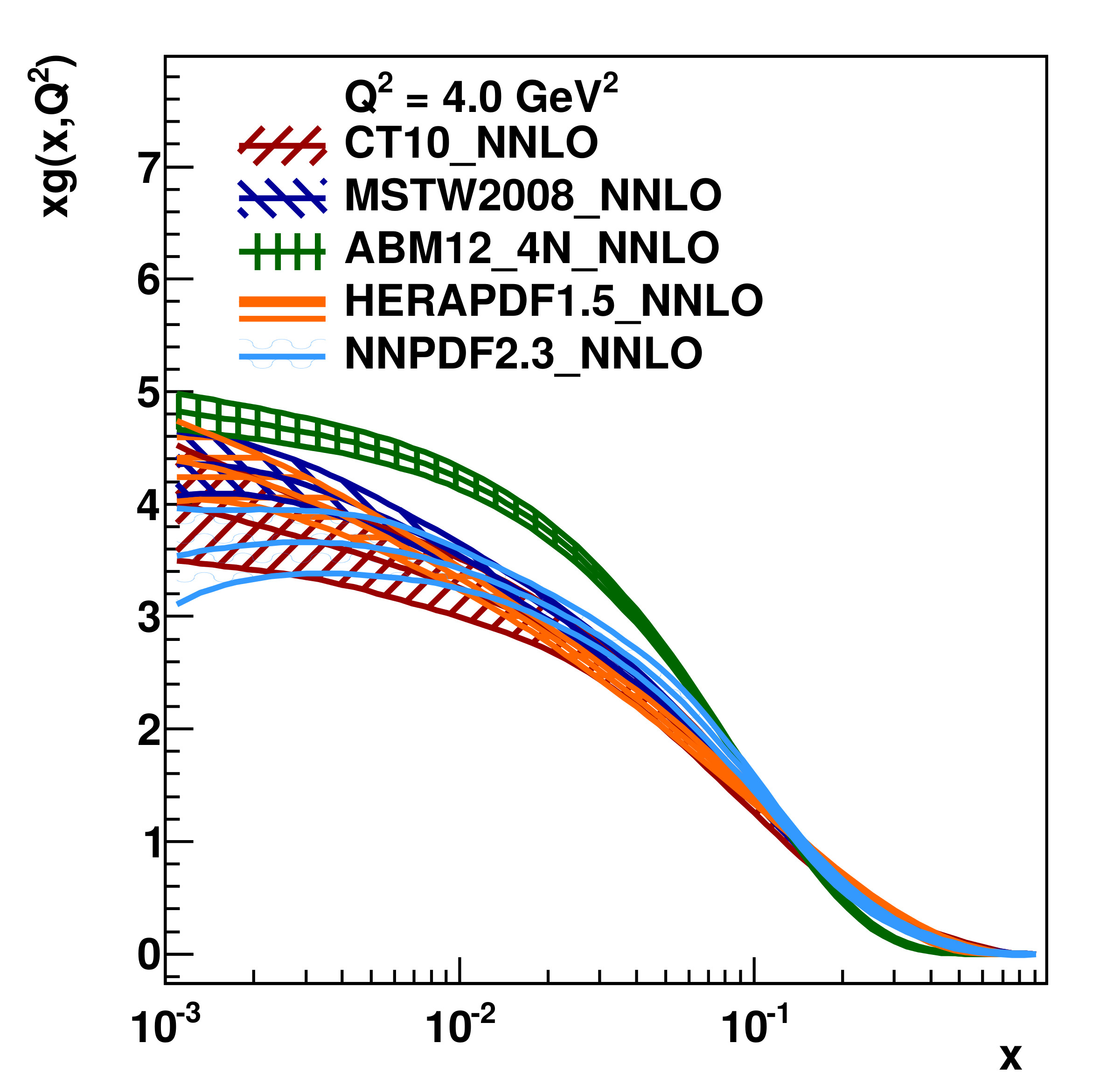}
    \caption{A selection of gluon PDFs with uncertainties at $Q^2=4.0\ {\rm GeV}^2$, 
plotted using the drawing tools from xFitter. \label{fig:pdf}
}
  \end{minipage}
  \hfill
  \begin{minipage}[t]{0.45\textwidth}
    \includegraphics[width=\textwidth]{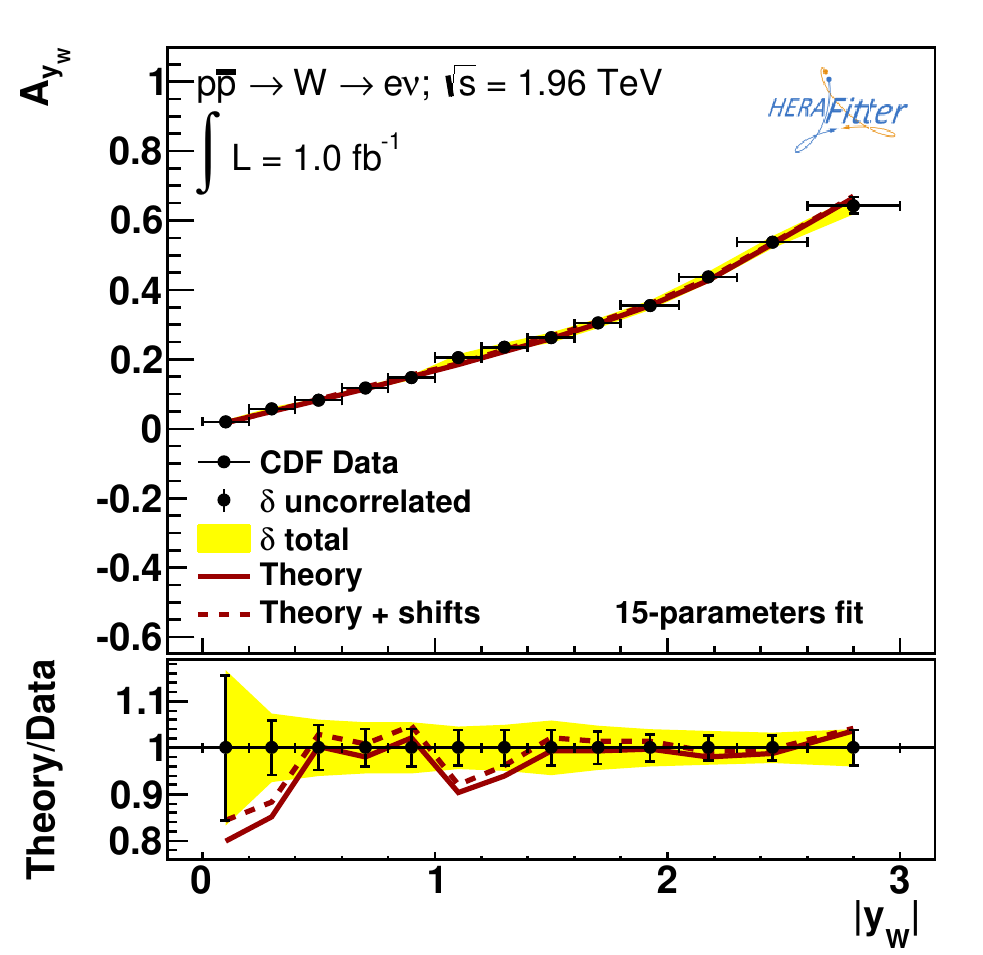}
    \caption{Comparison of theory predictions with Tevatron $W$ asymmetry  with uncertainties~\cite{Camarda:2015zba}. \label{fig:wz}
}
  \end{minipage}
\end{figure}
%%%%%%%%%%%%%%%%%%%%%%%%%%%%%%%%%%%%%%%%%%%%%%%%%%%%%%%%%%%%

The Parton Distribution Functions (PDFs) are the essential components
that allow us to make theoretical predictions for experimental
measurements of protons and hadrons.
The precision of the PDF analysis has advanced tremendously  in recent years, 
and these studies are now performed with very high  precision  at NLO and  NNLO in perturbation theory.

The xFitter project\footnote{%
xFitter can be downloaded from {\tt www.xFitter.org}.  An overview of
the program can be found in Ref.~\cite{Alekhin:2014irh}.  
} 
is an open source QCD fit framework 
that can perform PDF fits, assess the impact of new data, compare existing PDF sets, 
and perform a variety of other tasks~\cite{Alekhin:2014irh}.
The xFitter framework has already been used for more than 40 analyses
including many LHC studies.\footnote{%
A complete list is available at {\tt www.xFitter.org}.
}
The framework of xFitter is modular to allow for  various theoretical and
methodological options,  and contains interfaces to 
 QCDNUM~\cite{Botje:2010ay},
 APFEL~\cite{Bertone:2013vaa},
 LHAPDF~\cite{Buckley:2014ana},
 APPLGRID~\cite{Carli:2010rw},
 APFELGRID~\cite{Bertone:2016lga},
 FastNLO~\cite{fastnlo}  %{Britzger:2012bs,Stober:2015nlg},
 HATHOR~\cite{Aliev:2010zk}, % Kant:2014oha
among other packages. 
A schematic of the modular structure is illustrated in Fig.~\ref{fig:flow}.
xFitter also has a large number of data sets available, 
including a variety of fixed target experiments, 
HERA, Tevatron, and  LHC. 
It is also possible to add new custom data sets such as LHeC and EIC pseudo-data.

The xFitter project
grew out of PDF efforts of H1 and ZEUS which
became the HERA-Fitter project in 2012, and renamed the xFitter project
in 2015. xFitter is continually being updated, and version 2.0.0
(Frozen Frog) was released in March 2017 with many improvements and
new features.

In this short report we will provide a brief tour of selected xFitter
features with emphasis on some of the new additions;
we also direct the reader to the appropriate literature where
they can find more details.
Additionally, we will discuss some of the xFitter tutorials available
which can get the user up and running quickly with real analyses.

%%%%%%%%%%%%%%%%%%%%%%%%%%%%%%%%%%%%%%%%%%%%%%%%%%
\section{A Brief Tour}

The PDFs are the fundamental object that xFitter works with, and it
has a variety of utilities to read, write, and manipulate the PDFs and
associated uncertainties.
For example, xFitter is able to read and write PDFs in the LHAPDF6
format~\cite{Buckley:2014ana}, and Fig.~\ref{fig:pdf} illustrates a sample plot of the PDFs
from various global fitting projects.

xFitter can also generate comparison plots of data vs. theory, and an
example is shown in Fig.~\ref{fig:wz}.  There are a variety of options
for the definition of the $\chi^2$ function and the treatment of
experimental uncertainties.

\begin{figure}[!t]
  \centering
  \begin{minipage}[t]{0.45\textwidth}
    \includegraphics[width=\textwidth]{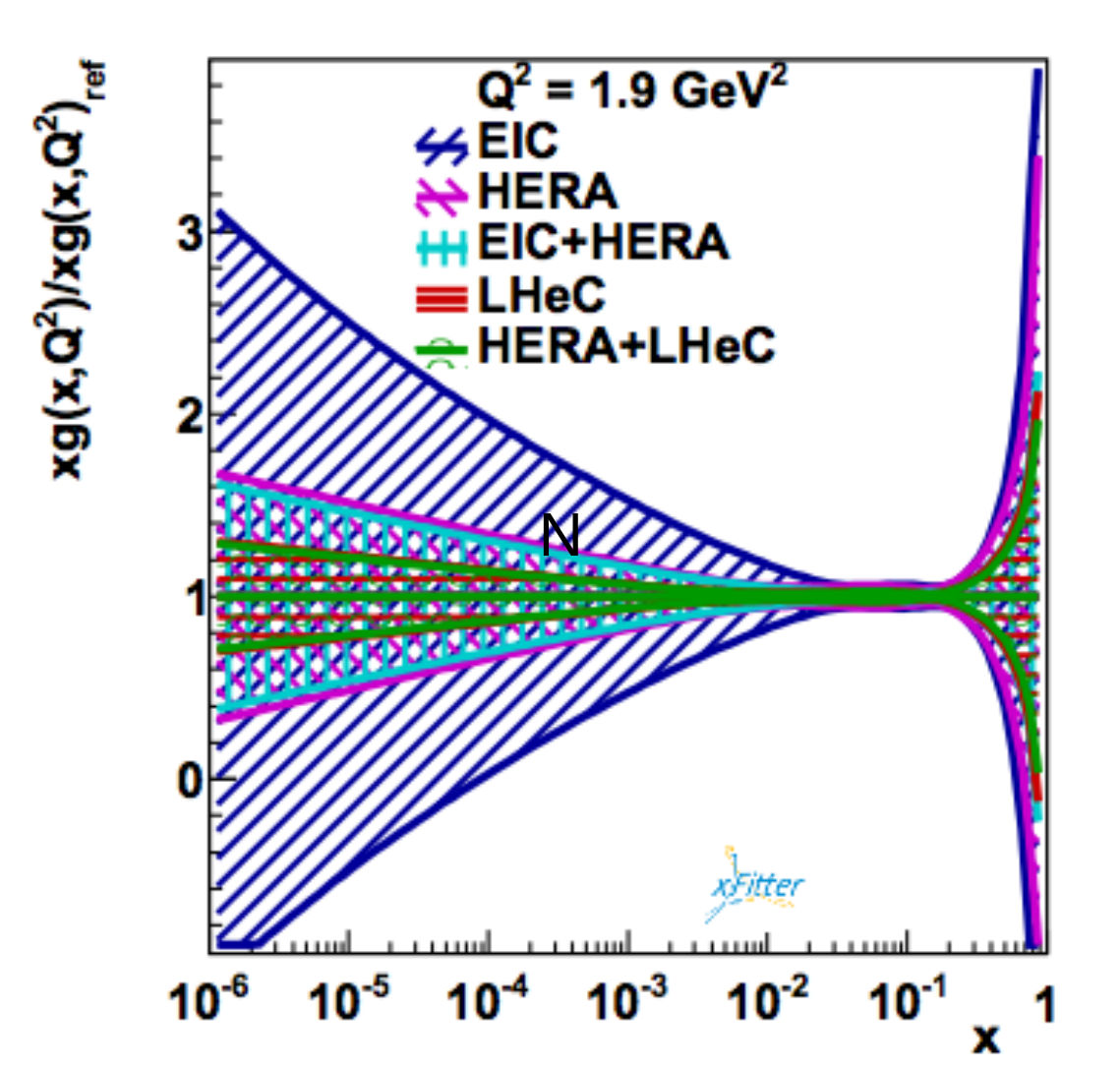}
    \caption{The PDF uncertainty as estimated by xFitter using pseudo-data  from different  
experiments~\cite{AbelleiraFernandez:2012cc}. \label{fig:lhec}
}
  \end{minipage}
  \hfill
  \begin{minipage}[t]{0.45\textwidth}
    \includegraphics[width=\textwidth]{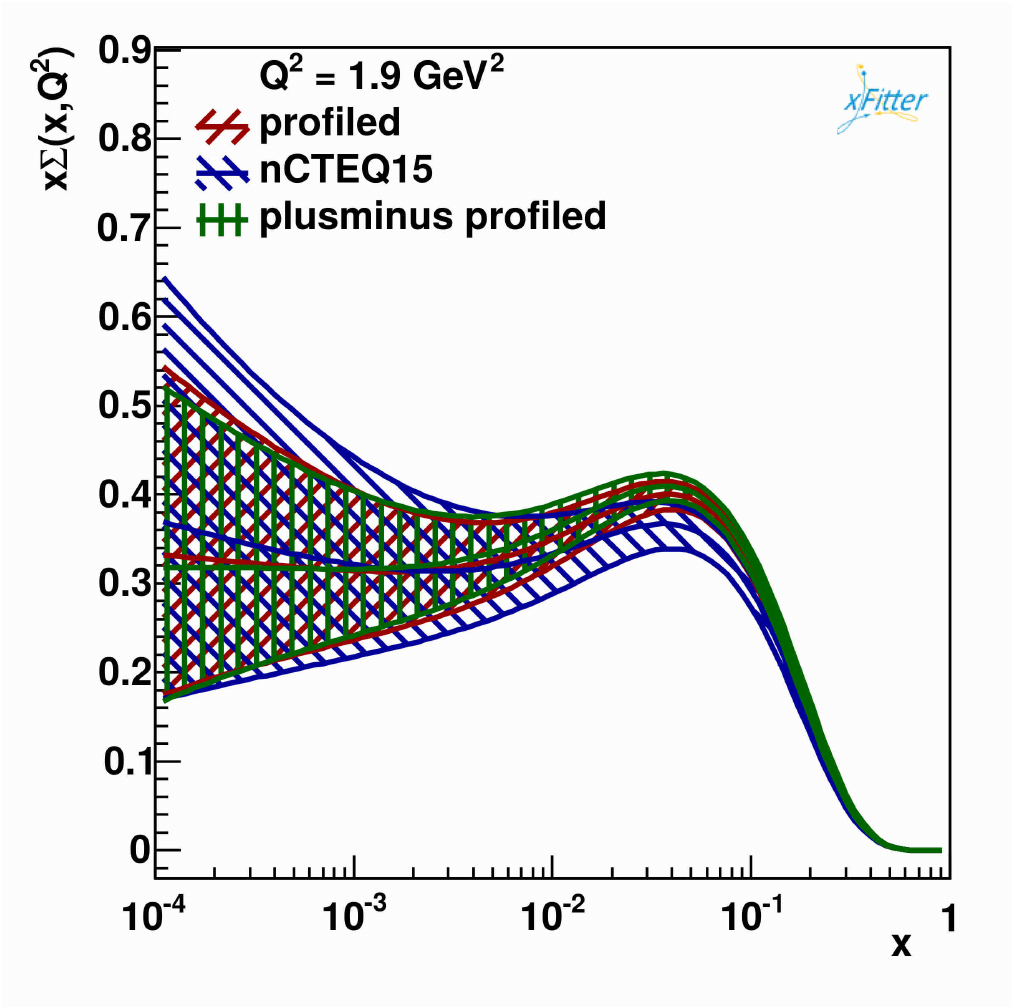}
    \caption{An example using xFitter profiling tools for nCTEQ15 lead nPDFs~\cite{Kovarik:2015cma}. \label{fig:lead}  }
  \end{minipage}
\end{figure}

An important application of xFitter is to understand how a particular
data set or experiment will impact the PDFs.  In Fig.~\ref{fig:lhec}
we show the results using pseudo-data from a proposed LHeC and EIC
experiment to constrain the relative uncertainty on the
gluon distribution.\footnote{%
The download of xFitter program includes a user manual and a sample
HERA data set.  Additional data sets are available at {\tt
  xfitter.hepforge.org} including the LHeC pseudo-data set.  }

%FIG 4,5
%%%%%%%%%%%%%%%%%%%%%%%%%%%%%%%%%%%%%%%%%%%%%%%%%%%%%%%%%%%%
%%%%%%%%%%%%%%%%%%%%%%%%%%%%%%%%%%%%%%%%%%%%%%%%%%%%%%%%%%%%

xFitter is able to perform PDF profiling and reweighting studies.
The reweighting method allows xFitter to update the probability
distribution of a PDF uncertainty set (such as a set of NNPDF
replicas) to reflect the influence of new data inputs.
For the PDF profiling, xFitter compares data and MC predictions 
based on the $\chi^2$-minimization, and then constrains the
individual PDF eigenvector sets taking into account the data
uncertainties.
We illustrate the profiling feature in Fig.~\ref{fig:lead} which
displays the impact of a new data set (W/Z production in LHC lead
collisions) on the nCTEQ15 PDFs~\cite{Kovarik:2015cma}.  In this case,
the additional data set has reduced the PDF uncertainty of the
original nCTEQ15 set.  This example also illustrates that while
xFitter cannot work with multiple nuclei simultaneously, it is able to
work with a single nucleus (in this case lead); hence, xFitter can also
be used for certain heavy ion analyses.

\begin{figure}[!t]
  \centering
  \begin{minipage}[t]{0.45\textwidth}
    \includegraphics[width=\textwidth]{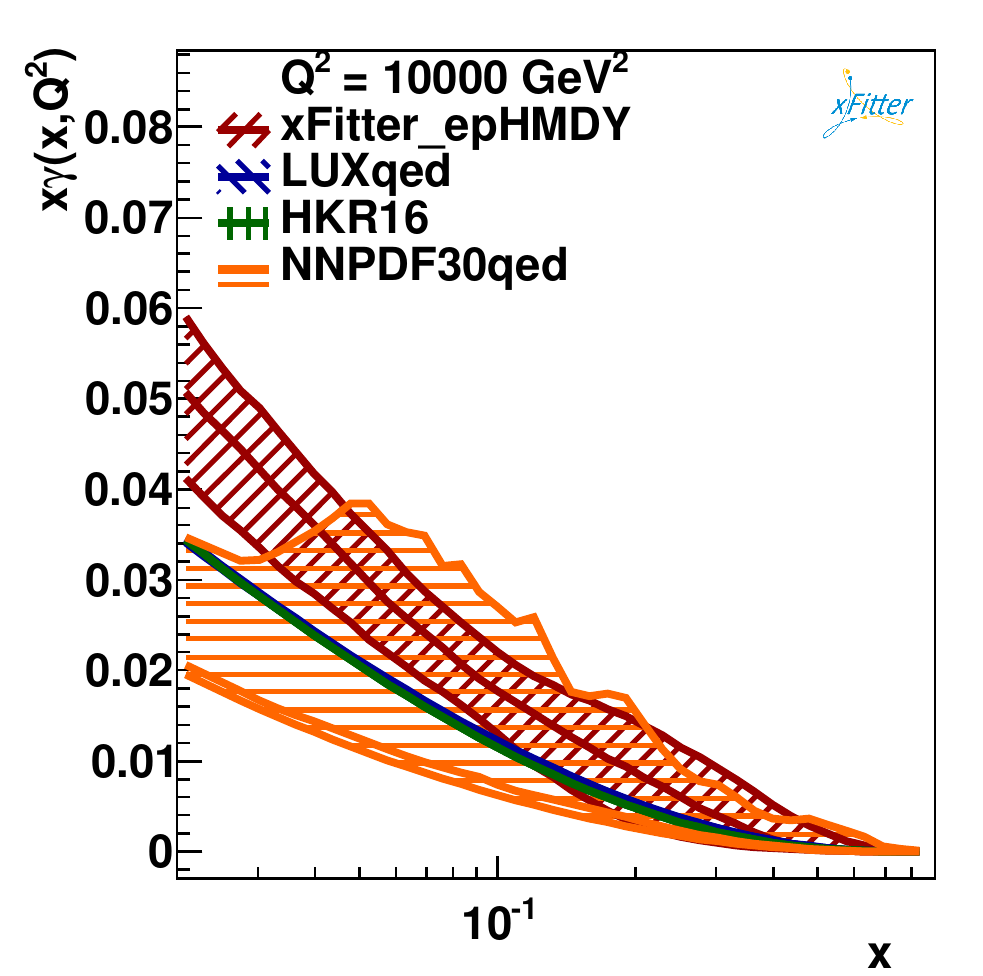}
    \caption{Comparison of the extracted photon PDF $x\gamma(x,Q^2)$ 
with other determinations from the literature~\cite{Giuli:2017oii}. \label{fig:photon}
}
  \end{minipage}
  \hfill
  \begin{minipage}[t]{0.45\textwidth}
    \includegraphics[width=\textwidth]{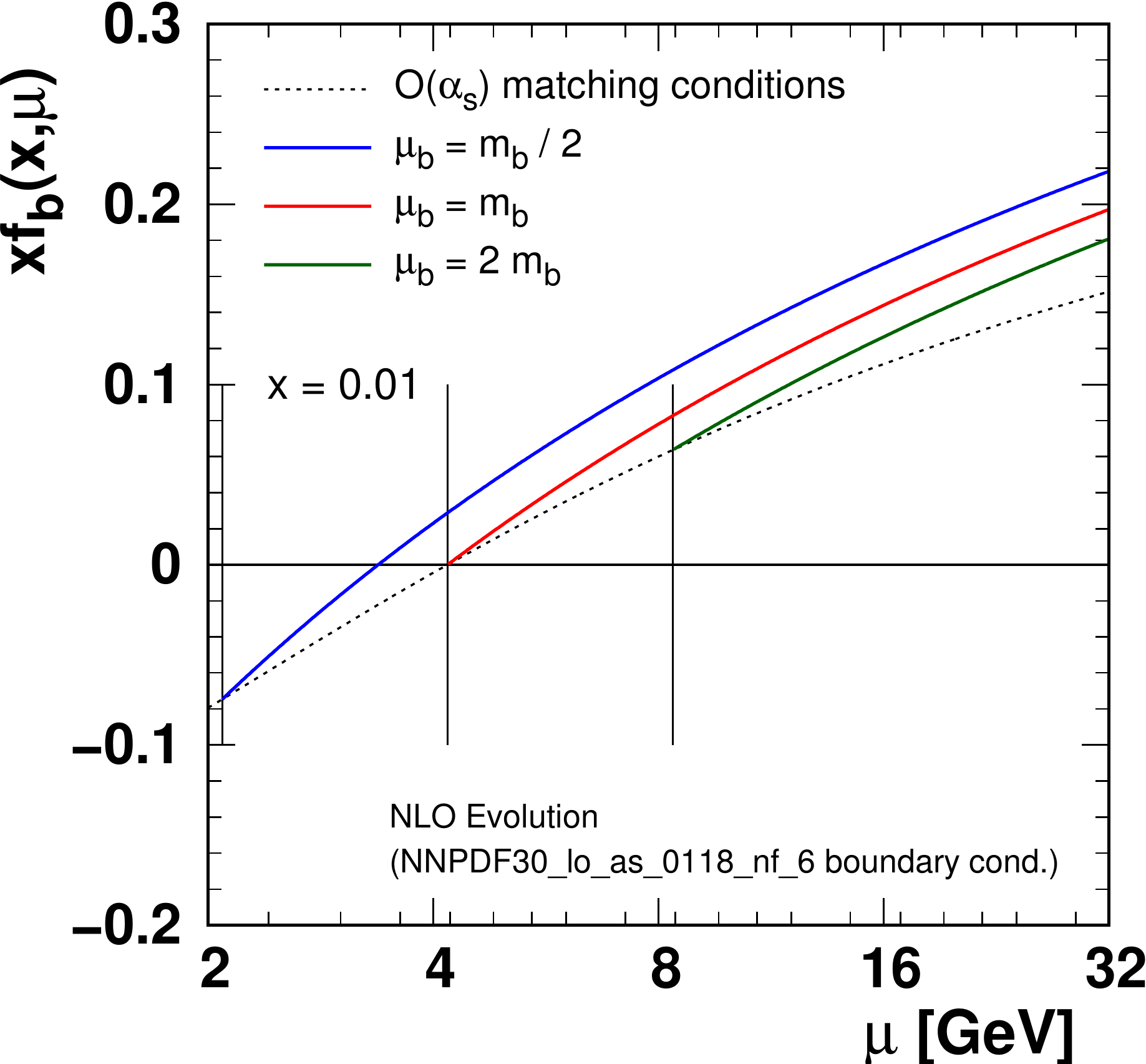}
    \caption{The $b$-quark PDF $xf_b(x,\mu)$ for different choices of the matching scales 
      $\mu_b=\{  m_b/2, m_b, 1 m_b \}$ as computed by xFitter
      in Ref.~\cite{valerio:2017tbd}. \label{fig:heavyq}
}
  \end{minipage}
\end{figure}

As many PDF analyses are now extended out to NNLO 
[{\cal  O}($\alpha_S^2$)], the NLO QED effects [{\cal O}($\alpha^1$)]
can also become important.\footnote{%
See presentation by F.~Giuli in these proceedings, ``The photon PDF
from high-mass Drell-Yan data at the LHC.''  }
For example, including QED processes in the parton evolution will
break the isospin symmetry as the up and down quarks have different
couplings to the photon.
xFitter is able to include NLO QED effects, and this is illustrated in
Fig.~\ref{fig:photon} which displays the photon PDF as determined
using a NNLO QCD and NLO QED analysis~\cite{Giuli:2017oii}.

Another new feature of xFitter is the ability to handle both pole
masses and $\overline{MS}$ running masses in the FONNL scheme. 
While the pole mass is more closely connected to what is measured in
experiments, the $\overline{MS}$ mass has advantages on the
theoretical side of improved perturbative convergence.
xFitter was used to perform a high precision determination of the
$\overline{MS}$ charm mass in this new
framework~\cite{Bertone:2016ywq}.

%%%%%%%%%%%%%%%%%%%%%%%%%%%%%%%%%%%%%%%%%%%%%%%%%%
\section{Available Tutorials}

The best way to appreciate the capabilities of the xFitter program is to 
try the program with a variety of exercises. 
Here we provide a brief overview of some of the available  tutorials. 
This range of applications  should be sufficiently comprehensive 
to serve as a starting point for many types of xFitter analyses.\footnote{%
The tutorials can be downloaded from the main xFitter website: {\tt www.xFitter.org}.
Examples~1-5 were designed by Stefano Camarda and presented at the ``2016 CTEQ-MCnet School''  at DESY,
and Example~6 was designed by Voica  Radescu and presented at the 
``Proton Structure in the LHC Era - School and Workshop.''
}

%FIG 6,7
%%%%%%%%%%%%%%%%%%%%%%%%%%%%%%%%%%%%%%%%%%%%%%%%%%%%%%%%%%%%
%%%%%%%%%%%%%%%%%%%%%%%%%%%%%%%%%%%%%%%%%%%%%%%%%%%%%%%%%%%%

%\begin{itemize}

% \item 
{\bf Example 1:} This is a  basic PDF fit using the HERA I+II
  data just to become familiar with the separate xFitter inputs,
  outputs, and drawing tools.  xFitter can compute at LO, NLO, and
  NNLO, and this exercise demonstrates a NNLO fit.

% \item 
{\bf Example 2:} This tutorial builds upon the previous exercise
  by using the HERA jets data to fit both the PDF and $\alpha_S$. In
  xFitter, $\alpha_S$ can either be fixed or floating, and we
  demonstrate this feature. Additionally, the HERA jet data has
  uncertainty correlation matrices, and we illustrate how xFitter can
  work with a variety of statistical and systematic uncertainties.

% \item 
{\bf Example 3:} This exercise illustrates the profiling methodology.
Starting with a given PDF in LHAPDF format, we add the Tevatron W asymmetry data set
(with  correlations) to determine  the impact on the initial PDF. 
xFitter will then output the modified PDF set in to a LHAPDF format 
which can be used for calculation or comparison. 

% \item 
{\bf Example 4:} There are variety of  tools which are useful 
to examine xFitter output. This exercise uses  Python with a Jupyter notebook
to read the LHAPDF output from the previous xFitter example, manipulate the results,
and generate comparison plots. Additionally, other tools are also available such
as the Mathematica  based ManeParse program~\cite{Clark:2016jgm}.

% \item 
{\bf Example 5:} 
xFitter has a variety of tools available for the $\chi^2$ definition
and the  treatment of experimental uncertainties. 
This exercise demonstrates the equivalence of using the covariance matrix representation and the 
nuisance parameters (correlated systematic  uncertainties) by running parallel fits and comparing 
the output.

% \item 
{\bf Example 6:} 
The measurement of the strange to down-sea quark ratio $r_s=s(x)/\bar{d}(x)$
has generated significant interest in  the literature~\cite{Aad:2012sb}
as the result was near the SU(3) limit $r_s\sim 1$. 
This measurement uses LHC $W/Z$ production data to extract the strange PDF.
This analysis can be reproduced in xFitter to allow the user to vary 
the inputs and the fitting parameters to validate the stability of this result. 

%\end{itemize}

While  the above examples provide only a brief glimpse of xFitter's capabilities, 
the breadth of these topics should enable the users to quickly gain the  necessary 
experience to proceed to their individual project.

%%%%%%%%%%%%%%%%%%%%%%%%%%%%%%%%%%%%%%%%%%%%%%%%%%
\newpage
\null \vspace{-1.4cm}
\section{Conclusion}

The xFitter 2.0.0 program is a versatile, flexible, modular, and
comprehensive tool that can facilitate analyses of the experimental
data and theoretical calculations.
xFitter has been used for a variety of analyses of fixed target,
Tevatron, HERA, and LHC data.
It is a valuable tool for bench-marking and understanding differences
between PDF fits, and it can provide impact studies for possible
future colliders such as the LHeC, EIC, and FCC.
We encourage use of xFitter, and welcome new contributions from the
community to ensure xFitter continues to incorporate the latest
theoretical advances and precision experimental data.

%%%%%%%%%%%%%%%%%%%%%%%%%%%%%%%%%%%%%%%%%%%%%%%%%%
%\section{Acknowledgment}

%Giuli:2017oii,Alekhin:2014irh,Camarda:2015zba,Kovarik:2015cma}
% Clark:2016jgm,Alekhin:2011sk}

%\lipsum{1-20}

%\section{Conclusion}

%%%%%%%%%%%%%%%%%%%%%%%%%%%%%%%%%%%%%%%%%%
%\bibliographystyle{plain}
%\bibliography{refs}

%\begin{thebibliography}{99}
%\bibitem{...}
%....
%\end{thebibliography}

%
% \newpage
%\bibliographystyle{spphys}
%\bibliographystyle{plain}
%\bibliographystyle{abbrv}
%\setlength{\bibsep}{0pt plus 11.3ex}
%\begin{spacing}{1.0}
% \bibliographystyle{srt}
% \bibliography{refs}
%\end{spacing}

%%%%%%%%%%%%%%%%%%%%%%%%%%%%%%%%%%%%%%%%%%
% \end{document}
%%%%%%%%%%%%%%%%%%%%%%%%%%%%%%%%%%%%%%%%%%

\null \vspace{-0.44cm}
\begin{spacing}{1.0}

\end{spacing}

%%%%%%%%%%%%%%%%%%%%%%%%%%%%%%%%%%%%%%%%%%
\end{document}